International Conference of Sabaragamuwa University of Sri Lanka 2015 (ICSUSL 2015)

# Evaluation of the use of web technology by government of Sri Lanka to ensure food security for its citizens

P. C. Abeysiriwardana[a,*], S. R. Kodituwakku[b]

[a]*Ministry of Science, Technology and Research, Galle Road, Colombo 3, 00300, Sri Lanka*
[b]*Department of Statistics and Computer Science, Faculty of Science, University of Peradeniya, Peradeniya, 20400, Sri Lanka*

**Abstract**

Web technology is one of the key areas in information and communication technology to be used as a powerful tool in ensuring food security which is one of the main issues in Sri Lanka. Web technology involves in communicating and sharing resources in network of computers all over the world. Main focus of food security is to ensure that all people have fair access to sufficient and quality food without endangering the future supply of the same food. In this context, web sites play a vital and major role in achieving food security in Sri Lanka. In this case study, websites pertaining to Sri Lankan government and link with food security were analyzed to find out their impact in achieving the goals of food security using web technologies and how they are being involved in ensuring food security in Sri Lanka. The other objective of this study is to make the Sri Lankan government aware of present situation of those websites in addressing food security related issues and how modern web technologies could be effectively and efficiently used to address those issues. So, the relevant websites were checked against several criteria and scores were used to assess their capabilities to address the concerns of food security. It was found that the amount of emphasis given by these websites to address the issues of food security is not satisfactory. Further, it showed that if these web sites could be improved further, they would generate a powerful impact on ensuring food security in Sri Lanka.

* Corresponding author. Tel.: +94-071-867-4258
  *E-mail address:* abeysiriwardana@yahoo.com

# 1. Introduction

*1.1. Government administration and services*

Sri Lanka is a multicultural country with many food traditions and a population of 20.35 million as at 2012 census[1]. For administrative purposes, Sri Lanka is divided into nine provinces and twentyfive districts. The districts are further subdivided into divisional secretariats and divisions which are attached to the divisional secretariats. There are 3 types of governments in Sri Lanka. District and divisional level administration is done by Central government. Provincial level administration is done by Provincial government linked to central government. The third and lowest level administration which links to Central government is done by three other types of local authorities called Municipal Councils, Urban councils and Pradeshiya Sabha. All of these administrative units have their own government institutes to provide public service free of charge or at discounted rate. Sri Lanka is one of the few countries in the world that provides universal free education from primary to tertiary stage[2] and free universal healthcare[3]. In addition to these services government has control on several services in relation to food and food security. Government has institutional capacity to handle food production, food distribution with a mechanism to look after fair distribution of food and ensure good practices on food related issues. Despites these services 22% of whole population i.e. 4.7 million people of Sri Lanka are undernourished according to The FAO Hunger Map 2015[4].

*1.2. Data on government web*

The data pertaining to food, in relation to the whole population and dispersed throughout these institutes hasn't been centralized or decentralized properly. Most of data in paper format or databases are in isolated computers or computer systems. But government has taken initiatives to give more value to these food related data by putting most of these data on internet and web to provide more effective and efficient food related services to public. This in turn positively contributes in ensuring food security in Sri Lanka.

Government Information Center (GIC) of Sri Lanka plays vital role in controlling and managing information about 1439 Services offered by 203 Organizations. GovSMS links with Government Information Center to obtain about 13 services from Government Departments via Short Messaging Services (SMS). GovSMS platform has SMS short code 1919, which is unified across all GSM and CDMA operators[5]. According to the government web portal, as at 15th April 2015, there are 650 websites pertaining to Ministries, Statutory Bodies and cooperation, Provincial Councils, District Secretariats, Divisional Secretariats, Departments, Government owned companies[6, 7]. Most of the websites used two national/official languages namely Sinhala, Tamil[8] and the link language, English to provide their services. Among these web sites, 82 web sites directly or indirectly provide the information about food to citizens of Sri Lanka. In addition to that, GIC through its short code 1919 and specially Department of Agriculture, Sri Lanka, through its short code 1920 provide information in relation to food to the public[9].

*1.3. Use of technologies to disseminate the information on food security*

Sri Lanka is the first country in the South Asian region to introduce 3G, 3.5G HSDPA, 3.75G HSUPA and 4G LTE mobile broadband Internet technologies[10] enabling these web services more accessible to its citizens.

Still with this advancement in the computer science, Sri Lanka is far away from using new technologies effectively and efficiently in its public service through government websites. Although Sri Lanka advanced to 65th place in 2015 from 76th in 2014 of Network readiness index[11], it doesn't seem government owned websites sufficiently use its power of web technologies to tackle issues of food security.

*1.4. Importance of web-technologies in food security*

The United Nations Food and Agriculture Organization (FAO) defines food security as "A situation that exists when all people, at all times, have physical, social and economic access to sufficient, safe and nutritious food that meets their dietary needs and food preferences for an active and healthy life."[12]. According to this definition, data/ information on four food security dimensions can be identified as follows.

1. Data/ information on food availability
2. Data/ information on economic and physical access to food
3. Data/ information on food utilization
4. Data/ information on food stability over time.

One of effective way of handling huge data pertaining to food security is to use of new technologies like web technologies. Because of the rapidly increasing use of mobile phones in developing countries, it has become another major breakthroughin dissemination of information. So, mobile 'OK' web sites will abundantly be used as a method of handling these data.

Today human needs, specially, in food related areas to be satisfied by the government in a post industrial era, are so vast that information and services needed by its citizens become very much complex. Web-technology is one such technology that can be used to satisfy these citizens' complex requirements effectively and efficiently. The requirement of handling these data is very important in the view of the concerns over food security which have been prompted by the impact of climate change, sharp rises in food prices, and energy policies in particular the issue of biofuels whose production requires the use of agricultural land. While handling these data/ information through web, there are other web related services which could be used to help ensuring food securities.

1. E-agriculture services can help farmers/ producers by posting advice direct to web forums, as well as weather forecasts and market prices.
2. Producers can go online to learn about new techniques – E-learning systems.
3. Discussion forums to help farmers, agriculturists in remote areas to improve their farming methods and products
4. Advertising on web to trade and find new markets and partnerships for farmers and producers. This will also help to avoid food being wasted, too.
5. Centralized databases on web can facilitate mapping agricultural production and food shortages by analyzing existing and projected data/ information.
6. Early-warning systems incorporated to web can help save food disasters as well as lives.

In rural areas, information can also be made freely and cost effectively available through web sites to farmers by community Telecenters like Nanasala, Vidatha, Cyber centers of Department of Agriculture which provide access to the Internet, telephone and fax services for rural or under served populations.

So, it is very important to analyze the government effort to initiate the use of web technologies to ensure food security, and how effective has been its use of this technology in addressing food security related issues. The other objective of this study is to make the Sri Lankan government aware of present situation of those websites in addressing food security related issues and how modern web technologies could be effectively and efficiently used to address those issues. In this survey government controlled 82 websites were tested to evaluate whether they are equipped with proper web technologies as well as information and applications in web to provide web based services regarding food security to public. For this, a self-questionnaire was developed and applied to each web site to evaluate for its web capacity to tackle food security.

## 2. Materials and methods

*2.1. Materials*

In this study, 82 government controlled/owned websites relevant and link to Ministries, Statutory Bodies and Cooperations, Provincial Councils, District Secretariats, Divisional Secretariats, Departments and Government owned companies were inspected during the months of April, May and June 2015.

The 82 websites selected for this study were among the 650 websites listed under the Lanka Gate – the Country Portal (https://www.srilanka.lk) which is a gateway for citizens to access information and services provided by the Government of Sri Lanka. All web sites are under same roof with so many similarities which could at the same time be considered as an advantage to improve aspects of food security by bringing interrelated subjects relevant to food security under same domain.

Some simple validator software[13] to test components which affect capacity of a web to provide effective services related to food security is also used.

*2.2. Methodologyn*

| Criteria no. (i) | Criteria | Value (V) | Weight (W) |
|---|---|---|---|
| 1 | Presence of the statements which indicates some commitment for any food security service in the objectives of the web site | 0 or 1 | 0.2 |
| 2 | Presence of the applications in the website which provides analytical power of data to user regarding food security | | |
| | 2.1 Self created | 0 or 1 | 1.4 |
| | 2.2 Third party | 0 or 1 | 1.1 |
| 3 | Presence of the other documentation provided by the web site which states its commitment for food security | 0 or 1 | 0.1 |
| 4 | Presence of interactive applications on the website to provided special services regarding food security. Ex: advertising, marketing, education, advising etc. relevant to food security to user | 0 or 1 | 1.5 |
| 5 | By using third party software to check standards of the website. | 0 or 1 | 0.2 |
| | 5.1 W3C mobileOK Checker v1.4.2 - A Web Page is mobileOK when it passes all the tests[16] | 0 or 1 | 0.1 |
| | 5.2 Feed Validator - Checks news feeds in formats like ATOM and RSS[17]. | | |
| 6 | By checking how the website stores and manages the incoming and outgoing data relevant to food security. | 0 or 1 | 0.3 |
| | | 0 or 1 | 0.1 |
| | 6.1 store its data on food security in a standard structure (database – SQL, Oracle, MySQL etc.). 6.2 handle in non-standard way (emails, text documents etc.) | | |

First, all 650 websites listed under government web portal were checked to select 82 websites which could make some connections with food security either by knowledge dissemination or by providing interactive applications which facilitate problems solving related to food security. Then those 82 sites were checked qualitatively to evaluate the amount ofcontribution of the website to be lined with food security related activities.

Checking the website for its capacity to tackle the issues regarding food security is a kind of quality check of that website. Although it is not totally a quality check it checks an existing functionality associated with quality. Most of the existing evaluation methods assess the accessibility and the usability of a website as indictors to test quality and performance of websites[14]. Usability test can be used to evaluate a website for its capacity of user computer interaction[15]. So, the interactive nature applications regarding food security is also looked for. There is no such critical literature review or description regarding a survey or study regarding evaluation of web-technologies for food security by providing such services to public through government intervention. Therefore, here only qualitative assessment is developed as an initiative to a quantitative assessment.

*2.3. Questionnaires, assumptions and criteria*

The criteria mentioned in the Table 1 were made to check the capacity or quality of the website in handling food security related issues by websites of government of Sri Lanka. The criteria emphasis on key areas or functionalities with respect to food security related websites and may directly support public interest in some instances. If the website is equipped with more criteria it becomes more effective in addressing food security related issues.

Table 1. Criteria to measure web technology capacity of government website to serve the purpose of food security.

The following two assumptions were made.

Assumptions:

1. The content in the government made or owned web site has no harmful or dubious or untruth information as they are controlled by government officials and standards of good governance.
2. When a website allocates space to talk about food, website is in a position to do something or show its concern on food security directly or indirectly. But the presence of such space doesn't necessarily indicate website has already acted on the thing.

*2.4. How to calculate web technology capacity on food security*

Value column in Table 1 denotes the presence or non-presence status of a criteria (which has no sub criteria) and sub criteria (when a main criteria has sub criteria). The set of government web sites is not very competitive in providing services with high impact to address food security issues. So, some simple criteria were used to categorize the capacity of these government web sites. Accordingly, each and every websites were inspected to find out whether it supports or doesn't support the criteria/ sub criteria by giving value 1 or 0 against each and every criterion/ sub criterion with respect to the inspected website. A criterion with sub criteria supposed to be satisfied when at least one of its sub criteria was satisfied. Weight is for measuring the impact of a criteria/ sub criteria on food security related subjects. Once the criteria value set was completed it was multiplied by Weight value set mentioned in Table 1. All criteria and sub criteria were considered to calculate the impact by website on food security related issues using <u>I</u>mpact <u>O</u>n<u>F</u>ood <u>S</u>ecurity index (IOFS).

Thus, IOFS was calculated by using Eq. (1),

$$IOFS = \Sigma V_i W_i \qquad (1)$$

Although the sample of websites used in this study is moderately large and selected from large set of web sites, the sample doesn't contain so highly competitive set of web sites to be evaluated for fine-tuned capacity of quality in the web site to address food security issues/ subjects. So the values for Weight were carefully assigned arbitrary, comparatively and relatively to each criterion to have fare impact of the criteria on capacity of web site to address food security issues/ subjects. Criteria are considered under two involvements in improving/contributing to food security as follows:

1. Content and application level – criteria 1, 2, 3 and 4
2. Technical involvement – criteria 5 and 6

If $2^{nd}$ or $4^{th}$ criteria or both are satisfied the website has more than good capacity to address food security issues/ subjects and IOFS value is > 1

If at least one of $2^{nd}$ or $4^{th}$ criteria is not satisfied but any combination of $1^{st}$, $3^{rd}$, $5^{th}$, $6^{th}$ criteria are satisfied website is considered as having a satisfactory to low capacity to address food security issues/ subjects and IOFS value is < 1 and >0.

Although the subject area of the website could have assisted in food security directly or indirectly but it doesn't provide any support in this regard will get 0 for IOFS, only if any of the criteria is not satisfied.

Based on the IOFS values, websites were given the qualitative judgment as mentioned in the Table 2, about its capacity to handle food security issues effectively and efficiently.

Table 2. Qualitative judgment on IOFS value.

| IOFS value | 0 | 0.1-0.4 | 0.5 – 1.0 | 1.1 – 2.4 | 2.5 – 5.0 |
|---|---|---|---|---|---|
| qualitative judgment | No support | Low | Satisfactory | Good | Very good |

It should be noted that these criteria can further improved by adding further sub criteria to them. For example how many statements, how many web pages, how many applications relevant to each criterion can be measured and associated with the IOFS for fine tuning of this value. But these websites are not so competitive against each other regarding food security, that kind of addition was not incorporated to IOFS at the present study.

## 3. Results, discussion, conclusion and recommendations

### 3.1. What is specially looked for

In any country, food system has evolved in size and complexity with global reach[18]. With this, food security plays a main role in state security with a natural trend to pay great attention also in national food security[19, 20]. There are so many factors which influence food security in any country which may list down as follows.

1. Population
2. Cultivated land
3. Food production/ yield
4. Total food demand
5. Quality of food
6. Security measures regarding food production/ food consumption
7. Storing and distribution of food
8. Sale of food and etc.

In this study, information regarding all of the above factors was considered. It is important to see how those data/ information originate, stored, processed, improved and used in different ways to tackle food security related issues. Effective and efficient use of these data/ information would contribute advancing of the science of food, ensuring a safe and abundant food supply, and healthier people everywhere. It essentially makes different and interesting connections to food securityrelated activities when those data/ information are in web.

Web technologies can be used to store, process, improve and use data/information in line with above factors effectively and efficiently in different ways to tackle food security related issues. When data and information on food security is on government web it allows a government's management to use data in a connected environment to look for food security needs, and demand trends to make effective strategic decisions.

*3.2. Results obtained and the interpretation*

In this study, 82 government web sites showed following results as depicted in Table 3 – 6 when tested against 6 main criteria and 6 sub criteria as mentioned in materials and methods section.

Table 3 shows the number of sites that satisfied each criterion used in IOFS. The criteria 1 which states the presence of commitment for food security in objectives of the web site, and the criteria 3 which states the presence of commitment in other web pages of the web site were met by more than 53% and 85% of sites, respectively. It could be seen that none of the site met the criteria of mobile OK. It should be noted that Even if Feed validator indicates the presence of Feed documents, it doesn't mean website has fully fledged feeds in support of food security. This is because of the quality of content in feed document of the web site is not guaranteed by the Feed validator. It only guarantees the validity of the feed in technical aspects.

Apart from mobile OK capacity, according to the Table 3, web sites have satisfactory level of technological capacity to handle food security issues. This is indicated by 34.15% of all web sites having feed validity, 43.90% sites have data on standard structures like databases and 75.61 % of all sites have ability to store/ transfer data using non-standard structures like email and text documents etc. But the ability to data analysis and providing interactive applications in these web sites is poor and reflected by low values of 9.76% and 3.66%, respectively. This showed that government websites lack real insight into the web technologies which can be used to serve people more effectively and efficiently in food security field.

Table 3. Number of web sites that satisfies each criteria.

| Criteria | 1 | 2 | | 3 | 4 | 5 | | 6 | |
|---|---|---|---|---|---|---|---|---|---|
| | | 2.1 | 2.2 | | | 5.1 | 5.2 | 6.1 | 6.2 |
| No. of Sites | 44 | 8 | 3 | 70 | 5 | 0 | 28 | 36 | 62 |
| Percentage of Sites | 53.66 | 9.76 | 3.66 | 85.37 | 6.10 | 0.00 | 34.15 | 43.90 | 75.61 |

It could be seen according to the Table 4 that more than one criterion had been satisfied simultaneously by most of the sites. Only 3 that is 3.66% sites had satisfied single criteria only while 6 sites (7.32%) didn't satisfied any criteria. 89.02% sites had satisfied more than one criteria simultaneously. 10 sites i.e. 12.20% of all sites could be listed as good or very good sites with respect to use of web technologies in food security.

Table 4. Number of web sites that satisfies each combination of criteria.

| | Criteria Present | Result | No. of websites | No. of websites % |
|---|---|---|---|---|
| 1 | 1 only | =< Satisfactory | 1 | 1.22 |
| 2 | 2 only (if any from 2 components satisfied) | Good | 0 | 0.00 |
| 3 | 3 only | =< Satisfactory | 1 | 1.22 |
| 4 | 4 only | Good | 0 | 0.00 |
| 5 | 5 only (if any from 2 components satisfied) | =< Satisfactory | 0 | 0.00 |
| 6 | 6 only (if any from 2 components satisfied) | =< Satisfactory | 1 | 1.22 |
| 7 | 2 AND 4 combination only | Very Good | 0 | 0.00 |
| 8 | Any combination with (2 OR 4) | Good → Very Good | 10 | 12.20 |
| 9 | Any combination excluding (2 OR 4) | =< Satisfactory | 63 | 76.83 |
| 10 | Sites which has no combination | No Commitment for Food Security | 6 | 7.32 |
| | Total | | 82 | 100 |

Table 5 shows 66 sites (80.49%) is either in satisfactory level or under it. Of these only 3 sites are with one criterion satisfied while rest of 63 sites has combination of criteria. Only 4.88% sites (4 sites) destined as very good web sites. They all are having combination of criteria. There are 6 sites (7.32%) categorized as good sites and they are also with more than one criteria satisfied for providing food security services.

Table 5. Summary of number of web sites that satisfy each category of IOFS.

| | With One criteria only | More than One criteria without analytical AND interactive applications | More than One criteria with analytical OR interactive applications | Total | sites % |
|---|---|---|---|---|---|
| Low - Satisfactory | 3 | 63 | 0 | 66 | 80.49 |
| Good | 0 | 0 | 6 | 6 | 7.32 |
| Very Good | 0 | 0 | 4 | 4 | 4.88 |
| No Commitment | 0 | 0 | 0 | 6 | 7.32 |
| Total | | | | 82 | 100 |

According to the Table 6, there are 39 web sites i.e. 47.56% of all web sites comes under category of low quality. This is because these sites have no proper technological involvement in web to provide food security services. These sites provided only some information/ data regarding food security to the users. 32.93% of all web sites (27 web sites) come under category of satisfactory level. This is because they have moderate technological capacity to be improved further for providing food security related services through web in addition of having some information/ data regarding food security to the users. Only 4.88% sites (4 sites) out of total sites went beyond the IOFS value of 2.5, thus, destined as very good web sites in this respect. These sites provide some kind of decision making, advising, analytical power or some similar functionality regarding food security to its web site users.

Table 6. Summary of number of web sites against IOFS values.

| qualitative judgment | No support | Low | Satisfactory | Good | Very good |
|---|---|---|---|---|---|
| IOFS value | 0 | 0.1-0.4 | 0.5 – 1.0 | 1.1 – 2.4 | 2.5 – 5.0 |
| No. of Sites | 6 | 39 | 27 | 6 | 4 |
| % sites against total sites | 7.32 | 47.56 | 32.93 | 7.32 | 4.88 |

*3.3. What government should look for in the direction of using web technologies for food security*

According to this study, it is apparent that although one third of government web sites reach satisfactory level only very few (12.20%) shows good to very good capacity to tackle food security concerns in modern world. Specially, when food and agriculture fields are considered, substantial amount of websites pertaining to them can be found. The amount of data pertaining to them is very large. The number of applications associated with them is also large. People interactions with them seem very high. But data were scattered between web pages and applications. If there were logical interactions between these application and data, the result would have been highly effective in solving citizens need. The amount of freely available third party software, interactive software and analytical software used in government web with respect to food security was not in satisfactory level.

The government should take advantage of the web which is in the process of redefining the meanings and processes of business, commerce, marketing, finance, publishing, education, research and development[21]. Although individual Web based information systems are moderately being deployed by government web sites (43.9%), advanced issues and techniques for developing and for benefiting from Web technologies for food security still remain to be a huge challenge.

Nevertheless, government already has a good web portal which could be used as an effective tool in improving web sites in relation to food security. So, the co-evolution of government web with food security will result in more user friendly and efficient government web service in the future in the field of food security to its citizens.

*3.4. Conclusion and recommendations*

In this paper, the importance of food security information/ data in modern web development was discussed. If government web sites can incorporated these food security related issues and information to its web sites in logical manner, government will be in good position to provide good government service with respect to food security to its citizens through its web sites.

In this study, 82 government web sites selected from 650 web sites listed under government web portal were checked for web technology capacity in solving food security related issues. It was found that more than one third of sites had capacity to reach the level of effective use of web technologies to help in addressing food security related issues because they had enough information and basic level of technologies to be thrived on more advanced technologies. But it was found that less than 15% sites had real commitment and advanced technologies to solve food security related issues. This shows that government should improve these web sites by incorporating advanced web technologies like web 2.0/3.0, third party software, interactive software and analytical software as well as incorporating the commitment to make this transition happen as the 21st century is the age of the Internet and the World Wide Web. Although individual web based information systems in line with food security related activities seem to be moderately deployed by government web sites, reaping the benefits by interconnecting the information systems still remains to be a huge challenge. If such connectivity could be established, it would provide one public web portal where citizens can access all services related to food security by single entrance.

So, it showed that use of web techniques open and publicly to provide web based services with respect to food security through government web to its citizens is not at satisfactory level. So, the lack of using the technologies pertaining to web in the food security hinders the most of the advantages that citizen and government can gain from such technological involvement with respect to food security related issues.